\documentclass{article}
\usepackage{frascatiphys}
\usepackage{graphicx}
\usepackage{rotating}
\usepackage{hyperref}
\usepackage{graphics}
\usepackage{color}
\usepackage{amssymb}
\usepackage{graphics, slashed}
\usepackage{amsmath}
\usepackage{slashed}
\usepackage{chngcntr}
\usepackage[normalem]{ulem} 
\usepackage {ulem} 
\usepackage{colordvi}
\usepackage{hyperref}

\def\s{\sqrt{s}}

\def\sigeff{\sigma_{eff}}
\def\siginel{\sigma_{inel}}
\def\sigtot{\sigma_{total}}
\def\siginelNSD{\sigma_{inel}^{NSD}}
\begin{document}
\title{
MODELING DOUBLE PARTON SCATTERING AT LHC}
\date{\today}
\author{Giulia  Pancheri \\
\em{INFN Frascati National Laboratories, 00044 Frascati, Italy}\\
Agnes  Grau\\
\em{Departamento de Fisica Teorica y del Cosmos, Universidad de Granada, 18071 Granada, Spain}\\
Simone Pacetti and Yogendra N. Srivastava  \\
\em{Department of Physics \& Geology, University of Perugia, 06123 Perugia, Italy}}
\maketitle
\baselineskip=10pt
\begin{abstract}
We examine present data  for double  parton scattering at LHC and discuss their  energy dependence  from its earliest measurements at the ISR.  
Different models for the effective cross-section are considered and 
their behavior studied for a variety of selected final states. 
We  point out that  data for $pp->4 \ jets$  or $pp\rightarrow \ quarkonium \ pair$ 
indicate $\sigma_{eff}$ to  increase with energy. 
We compare this set of data with different models, including one inspired  by our  soft gluon resummation model for the impact parameter distribution of partons.
\end{abstract}


\baselineskip=14pt
\section{Introduction}
Double parton scattering in hadron collisions has been searched for and measured for more than 30 years. 
Recently, the ATLAS collaboration \cite{Aaboud:2018tiq} has  examined  all existing data for Double Parton Scattering events, from  ISR  to LHC 13 TeV, and  a value for the the effective cross-section 
has been extracted. For a process of the type $p p \rightarrow A + B + X$ the following expression was used 
\begin{equation}
\sigma_{DPS}^{AB}=\frac{k}{2}\frac{\sigma^{A}_{SPS}{ \sigma^{B}_{SPS}}}
{\sigma_{eff}}
\end{equation}
with $k$ a symmetry factor to indicate identical or different final states, and  $\sigma_{eff}$  interpreted as the overlap area (in the transverse 
plane) between the interacting partons.\\
In this note,   the energy dependence of $\sigeff$ will be discussed in light of a few models and a rather general theorem. We shall start by presenting in Sect.~\ref{sec:general} the   general framework for multi parton scattering as recently presented by  D'Enterria in \cite{dEnterria:2017yhd}  and then apply this formalism to show that, in general, $\sigeff$ cannot be asymptotically a constant.  \\
 In Sect.~\ref{sec:BN} and Sect.~\ref{sec:ampli}  we consider  various   strategies for the calculation of $\sigeff$, a geometrical one in which  $\sigeff$ is interpreted  as the area  occupied by the partons involved in the collision and thus obtain it  from modelling the impact parameter distribution of partons, another one in which the area is directly obtained as the Fourier transform of the scattering amplitude. These different strategies may  lead to different energy dependence, as we shall see.
  \section{\bf Matter distribution in a hadron}
 \label{sec:general}
Theoretically multi-parton scattering (MPS)  has been of great interest  
\cite{Paver:1983hi,Ametller:1985tp,Sjostrand:1987su,Strikman:2010bg,Seymour:2013sya,Bartalini:2017jkk, Diehl:2017wew}. 
A key element in an analysis of an n-parton process (NPS) with final particle  states ($a_1,a_2,...a_n$) in terms of the single-parton
processes (SPS) is the role played by an {\it effective parton cross-section} defined as follows:
\begin{eqnarray}
\label{e1}
\sigma^{NPS}_{h_1h_2\to a_1,a_2,...a_n} = \Big{[} \frac{m}{\Gamma(n+1)} \frac{\sigma^{SPS}_{h_1h_2\to a_1}
\sigma^{SPS}_{h_1h_2\to a_2}.....\sigma^{SPS}_{h_1h_2\to a_n}}
{[\sigma_{eff,NPS}]^{n-1}}\Big{]}.
\end{eqnarray}  
 As Eq.~(\ref{e1})
deals with {\it probabilities} rather than {\it probability amplitudes}, it is clear that
the description is {\it semi-classical} and ignores any correlation between production of particles.
On the other hand, the degeneracy factor $m$ in these equations, to be defined momentarily,  does 
distinguish between identical and non-identical particle  states and thus must be thought as of quantum mechanical
origin. For a two  parton  process (DPS) (say, $a_1,a_2$), $m=1$ if the two particle  states are identical ($a_1=a_2$) and $m=2$ if 
they are different ($a_1\neq a_2$). For a three particle process (TPS), $m=1$ if $a_1= a_2= a_3$; $m=3$ if $a_1=a_2$ and
$m=6$ if $a_1 \neq a_2 \neq a_3$. Etc.\\
Under a set of {\it reasonable} hypothesis of factorization of parallel and transverse momenta, the quantity of interest $\sigma_{eff}^{NPS}$
is approximated in terms of the normalized {\it single} parton distribution  
or, generally a matter distribution
$T({\bf b})$ inside a hadron in impact parameter space, as follows
\begin{eqnarray}
\label{e3}
\int (d^2{\bf b}) T({\bf b}) =1;\ \ \
\Sigma^{(n)} \equiv \int (d^2{\bf b}) T^n({\bf b});\ \ 
\sigma_{eff}^{NPS} = [\Sigma^{(n)}]^{-1/(n-1)}
\end{eqnarray}
Before turning our attention to the  crucial input of  the single parton {\it overlap function}  
we present here an argument as to why $\sigeff$ cannot 
-in general
i.e., for  all types of final states in DPS or MPS scattering-  
be a constant.\\
In particular, we shall now show that if $\sigma_{eff}(s)$ approaches a constant as $s\to \infty$, then {\it all}, multi-parton cross-sections $\sigma^n_{a_1;....a_n}(s)$ must also approach constants asymptotically under the very mild hypothesis that 
$\sigma^{n+1}_{a_1;....a_{n+1}}(s) < \sigma^n_{a_1;....a_n}(s)$ for $a_i \neq a_j$.
Consider in fact  Eq.(9) of \cite{dEnterria:2017yhd}
\begin{eqnarray}
\label{1}
\sigma^{(2)}_{a_1;a_2}(s) = (\frac{m}2{})\frac{\sigma^{(1)}_{a_1}(s) \sigma^{(1)}_{a_2} (s)}{\sigma_{eff}(s)};
m =2\ {\rm if}\ a_1 \neq a_2; m=1\ {\rm if}\ a_1 = a_2;
\end{eqnarray}
in an obvious notation. Let 
\begin{equation}
\label{2}
(i)\ \sigma_{a_i}(s) \to L_i(s); {\rm where}\ L_i(s)\ {\rm increase\ with}\ s;\ \ \ \ \  (ii) \ \sigma_{eff}(s) \to {\rm a\ constant};\end{equation}
  Then, it follows from (i) and (ii)\ that
\begin{equation} {\rm for\ }    a_1 \neq a_2:  \sigma^{(2)}_{a_1;a_2}(s) \propto L_1(s) L_2(s) {\rm \ but\  then\ }\Rrightarrow 
\ \Big{[}\frac{\sigma^{(2)}_{a_1;a_2}(s)}{\sigma^{(1)}_{a_1}(s)}\Big{]} \propto L_2(s)\ {\rm increases\ with}\ s;
\end{equation}
and thus not bounded by a constant thereby violating the initial hypothesis. Hence, $L_2(s)$ can not increase with $s$ but must be bounded by a constant. We can repeat 
the proof by exchanging $a_1\leftrightarrow a_2$ and show that also $L_1(s)$ must be a constant. Ergo, also $\sigma^{(2)}_{a_1; a_2}(s)$ must go to a constant
as $s \to \infty$. \\
Extension of the above to the identical case ($a_1 = a_2$) and for $n=3,4,....$ are left as exercises to the reader. The proof is specially easy if Eqs.(3) \& (7) of \cite{dEnterria:2017yhd} are recalled.  In the next section,  we turn our attention to $T({\bf b})$. 

\section{The BN model for $\sigeff$}
\label{sec:BN}
 In this section we examine a model for $\sigeff$, in which    the impact parameter distribution of   partons is obtained from soft gluon resummation. As we shall see later,  this model reproduces the order of magnitude of $\sigeff$ but bears different energy trend depending on the PDF used. 
A suitable model for a normalized $T({\bf b})$ -albeit with a different name $A({\bf b})$- has been an object of our attention for over two decades and detailed references can be found in our review\cite{Pancheri:2016yel}. We start with a model  in which the area occupied by the partons involved in  parton scattering can be related to soft gluon resummation.  
In this model for the total cross-section, the energy behaviour of the total and inelastic cross-sections 
are obtained in the eikonal formalism, with {\it mini-jets}, partons with $p_t > p_{tmin} \approx 1.1-1.5 $ GeV, to drive the rise and soft gluon resummation to tame it. The impact parameter distribution is determined by the Fourier transform of the  $k_t$ distribution of  soft gluons emitted during semi-hard   parton scattering. Namely the normalized matter distribution in impact parameter space, $T({\bf b})$ in this model, is 
\begin{equation}
A(b,s)=N(s)\mathcal{F} [\Pi({\bf K}_t)]=N(s) \int d^2{\bf K}_t \int d^2{\bf b} e^{i{\bf K}_t \cdot {\bf b}}  e^{-h(b,s)}; \ \ \ \  h(b,s)=\int_0^{qmax} d^3{\bar n}(k)[1-e^{-i\bf{k}_t \cdot \bf{b}}] \label{eq:abs}
  \end{equation}
  where the overall distribution  $\Pi({\bf K}_t)$ is obtained by resummation of  soft gluons emitted with average number  ${\bar n}({\bf k})$. The above expressions are semi-classical and can be obtained by summing all the gluons emitted with momentum ${\bf k}_t$ in a Poisson like distribution. The effect of imposing energy-momentum conservation  to all  possible distributions  results in the factor among square brackets in Eq.~(\ref{eq:abs}). Such factor allows to integrate in $k_t$ down to zero, if ${\bar n}(k)$ is no more singular than an inverse power. While this is true in QED, for gluons this is not possible. In our model for the total cross-section, which is related to large distance behaviour of the interaction,   the impact parameter distribution is related to very small $k_t$ values. This implies including very small $k_t$ values, lower than $\Lambda_{QCD}$, values usually not included in the resummation or 
 ``lumped'' into an intrinsic transverse momentum. In order to evaluate $h(b,s)$ down to such low values, we proposed    a  phenomenological approximation for $\alpha_s(k_t\rightarrow 0)$, namely our phenomenological choice is
  \begin{equation}
  \alpha_s(k_t\rightarrow 0)\propto [\frac{k_t}{\Lambda_{QCD}}]^{-2p}; \ \ \ \ \ \ 
  \alpha_s(k_t >> QCD scale) = \alpha_s^{asym-free}(k_t)=\frac{1}{b_0 \ln{\frac{k_t^2}{\Lambda^2_{QCD}}}}
  \end{equation}     
with $1/2<p<1$. Our model for $\sigma_{eff}$, using  $A(b,s)$ from Eq.~(\ref{eq:abs}) is given as
\begin{equation}
\sigma_{eff}(s)=\frac{[\int d^2{\bf b} e^{-h(b,s)}]^2}{\int d^2{\bf b} e^{-2h(b,s)}}
\end{equation}
We have indicated that the function $h(b,s)$ depends 
upon the c.m.s energy of the collision, so will then be true also for $A(b,s)$. 
Because of the minimum transverse momentum  $p_{tmin}$ allowed to the minijet cross-section, $q_{max} $ will depend also on $p_{tmin}$. 
Through an  average procedure \cite{Godbole:2004kx}, one can obtain $<q_{max})>$ as a function of $\s,PDF,p_{tmin}$. The results from this resummation 
can then be used to  model the eikonal function and calculate inclusive quantities such as total and inelastic cross-section. In our model,  soft and semi-hard gluons contribute to the observed rise of the total cross-section with soft gluons  tempering  the fast rise (with energy) due to  the  mini-jet cross section. In Fig.~{\ref{fig:qmax-sigjet-tot} results for  $<q_{max}>$ and $\sigma_{jet}(\s,p_{tmin})$ are shown four  different LO PDFs, together with the total or  inelastic cross-section corresponding to the indicated  for parameter choice, including  updated PDFs, such as MSTW.
\begin{figure}
\hspace{-1.4cm}
\includegraphics[scale=0.4]{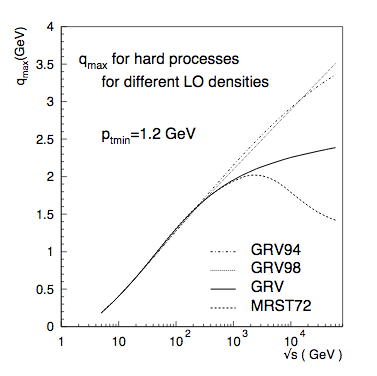}
\includegraphics[scale=0.42]{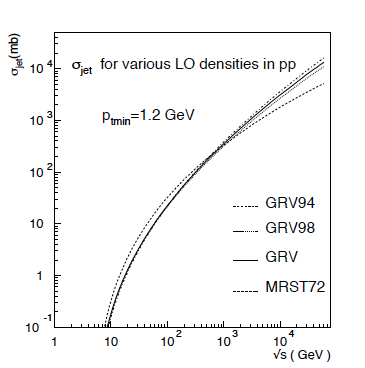}
\hspace{-.3cm}
\includegraphics[scale=0.258,angle=90]{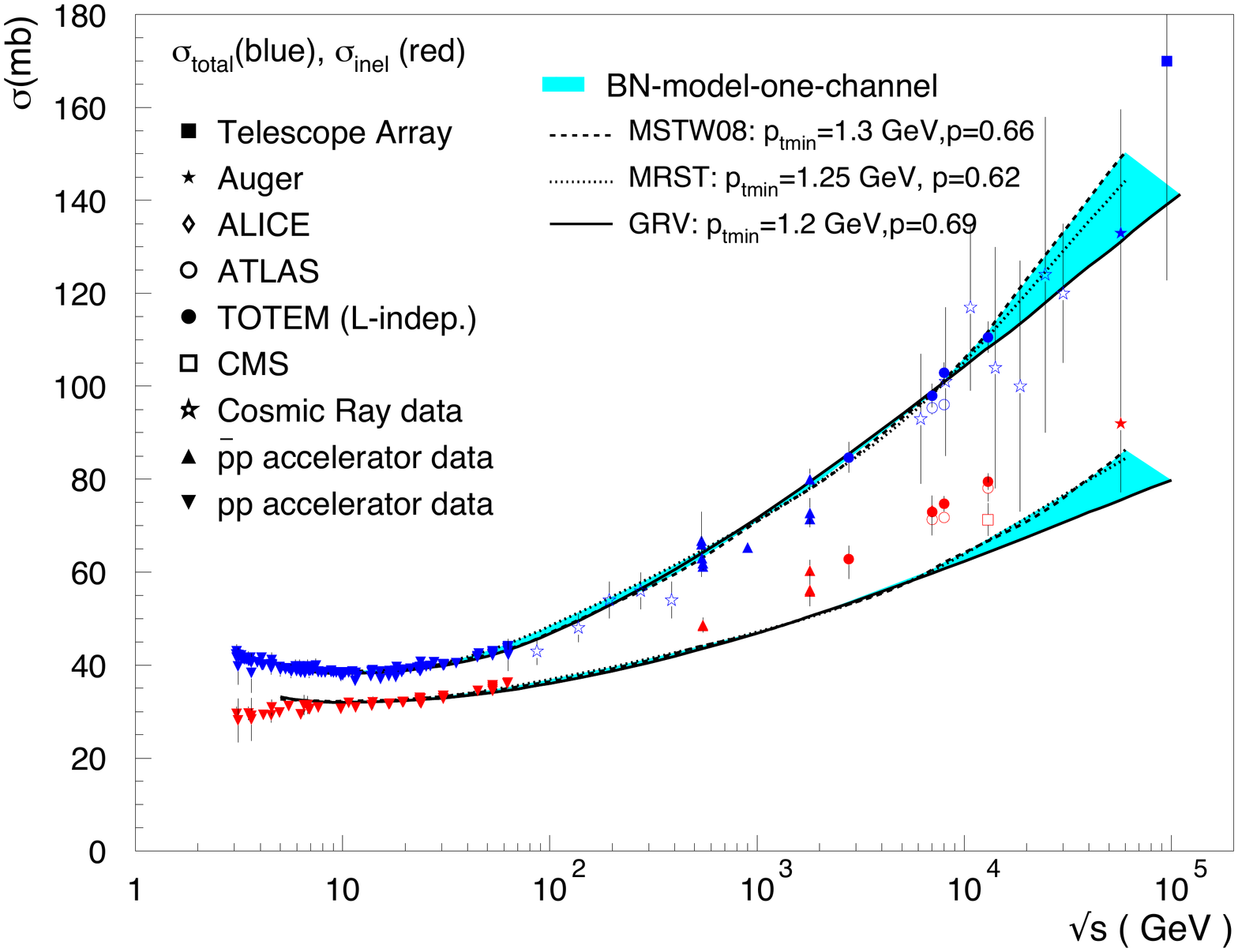}
\caption{For a given set of PDFs with corresponding $p_{tmin}$, left and center plots respectively show  the maximum transverse momentum allowed to soft gluons emitted by partons participating to a semi-hard collision and  the behaviour of the minijet cross-section used in the model from \cite{Fagundes:2015vba}. The  figure at  right shows  the corresponding description of total and Non Single Diffractive inelastic cross-section with appropriate   choices of the singularity parameter $p$.}
\label{fig:qmax-sigjet-tot}
\end{figure}
One should notice that the energy behaviour of $<q_{max}>$ is different  for different densities, as does  the one from $\sigma_{jet}$, but that they compensate in the predicted behaviour of the total cross-sections, which both smoothly rise in accordance with the Froissart bound, as shown in \cite{Grau:2009qx}. This will not be true for $\sigma_{eff}$, as 
 the model produces an energy dependence of $\sigeff$ which correlates only with the energy dependence of $q_{max}$, i.e. the upper limit of integration over soft gluon spectrum, so that if $q_{max} \uparrow \sqrt{s}$, then $\sigeff \downarrow \sqrt{s}$ and vice-versa.

\section{The elastic amplitude and $\sigeff$ energy dependence}
\label{sec:ampli}
According to \cite{dEnterria:2017yhd} and following the summary shown in Sec.~\ref{sec:general}, 
\begin{eqnarray}
\label{e3}
\int (d^2{\bf b}) T({\bf b}) =1;\ \
\Sigma^{(2)} \equiv \int (d^2{\bf b}) T^2({\bf b});\ \
\sigma_{eff}^{DPS} = \frac{1}{\Sigma^{(2)}}
\end{eqnarray}
The above arrives upon considering factorization between the hard jet cross-sections and the  impact parameter distribution of the involved partons, whose $\mathcal{F}$-transform gives the transverse momentum of partons involved in the hard cross-section. This model has a theoretical basis, but  one  needs an  expression for $T({\bf b})$ to use. 
The derivation in D'Enterria gives the following expression for T({\bf b}):
\begin{equation}
\label{e4}
T({\bf b})=\int d^2 {\bf b_1}f({\bf b_1})f(\bf{ b -  b_1})
\end{equation}
where $f({\bf b})$  describes the transverse parton density of the hadron.\\
Apart from phenomenological fits of the type $e^{-(b/scale)^m}$,  which have problems with analyticity if $m<1$ \cite{Grau:2009qx}
 consider what is at the root of the formalism being considered regarding the transverse spatial (in short, the b)-distribution adopted in Eq.(\ref{e4}). Also, we can recall the lessons learnt from analyticity of hadronic form factors and  the elastic amplitudes.\\
One begins with $f(\bf b)$, a normalized b-density function and its Fourier transform, the transverse momentum distribution $\hat{f}(\bf q)$ for a single parton, as follows:
\begin{equation}
\label{g1}
f({\bf b}) = \int[\frac{d^2{\bf q}}{(2\pi)^2}] e^{i {\bf b}\cdot {\bf q}}\ \hat{f}({\bf q});\ \ \  \hat{f}({\bf q}) = \int (d^2{\bf b}) e^{-i {\bf b}\cdot {\bf q}}\ f({\bf b});\ \ \ \hat{f}({\bf q=0}) =   \int (d^2{\bf b}) f({\bf b}) = 1
\end{equation}   
Let us consider  this parton 
 distribution first in momentum space and then in b-space. The simplest case to start with is that of collinear partons. 
The probability density that two-partons are at the same momentum transfer is given by 
\begin{equation}
\label{g2}
\hat{T}({\bf q}) \equiv\ [\hat{f}({\bf q}]^2; \ \ \ \ \ with \ \ \ \ \ \hat{T}({\bf q=0}) = 1
\end{equation}   
whose Fourier transform $T({\bf b})$ reads
\begin{equation}
\label{g3}
T({\bf b}) \equiv\  \int[\frac{d^2{\bf q}}{(2\pi)^2}] e^{i{\bf b}\cdot{\bf q}}\hat{T}({\bf q}) =  \int[\frac{d^2{\bf q}}{(2\pi)^2}] 
e^{i{\bf b}\cdot{\bf q}}[\hat{f}({\bf q}]^2
= \int (d^2 {\bf b}_1) f({\bf b}_1) f({\bf b} - {\bf b}_1)\\
\end{equation}   
which exactly reproduces Eq.(\ref{e4}). Also, by virtue of Eq.(\ref{g1};\ref{g2}), $T({\bf b})$ is properly normalized, viz.,
\begin{equation}
\label{g4}
 \int (d^2{\bf b}) T({\bf b}) = \int d^2({\bf b}) \int (d^2 {\bf b}_1) f({\bf b}_1) f({\bf b} -  {\bf b}_1)= [\int (d^2{\bf b}) f({\bf b})]^2  =1;\ \ \ 
\hat{T}({\bf q=0}) =  \int (d^2{\bf b}) T({\bf b})= 1. \end{equation}   
Now to some considerations about the effective cross-section $\sigma_{eff}(s)$, which for this simple identical parton model shall be taken to be (with a factor of a $1/2$)
\begin{eqnarray}
\label{g5}
2\sigma_{eff}(s) = \Big{[}\frac{1}{\int (d^2{\bf b})  T^2({\bf b})}\Big{]};
\end{eqnarray}   
but, by virtue of Eqs.(\ref{g2}) {\it et sec}, it follows that
\begin{equation}
\label{g6}
\int (d^2{\bf b})  T^2({\bf b}) = \int [\frac{d^2{\bf q}}{(2\pi)^2}] \hat{f}({\bf q}]^2 \hat{f}({-\bf q}]^2
=  \int [\frac{d^2{\bf q}}{(2\pi)^2}] |\hat{f}({\bf q}|]^4;\nonumber\\
\end{equation}   
Since, $\hat{f}({\bf q=0}) =1$, at first sight, it may appear reasonable to assume that it is the elastic form factor. So, for this form factor assuming the dipole form, we have
\begin{eqnarray}
\label{g7}
\hat{f}({\bf q}) = \frac{1}{[1 + (q^2/t_o(s))]^2 };\nonumber\\
\sigma_{eff}^{(el)}(s) =\frac{1}{ \int[\frac{d^2{\bf q}}{(2\pi)^2}] \frac{1}{[1 + (q^2/t_o(s))]^8}}
= \Big{[} \frac{14\pi}{t_o(s)} \Big{]}.
\end{eqnarray}
To get a simple estimate, we can employ  the result from a fit to the elastic differential cross-section, discussed in \cite{Fagundes:2013aja}. At 13 TeV, our estimate for  the  elastic scattering form-factor value (work in preparation)}  is 
 $t_o(13\ TeV) \approx\ 0.6\ GeV^2$, leading to
\begin{eqnarray}
\label{g8}
\sigma_{eff}^{(el)}(13\ TeV) \approx\ 28.6\ milli-barns.
\end{eqnarray}   
We notice that
the  value predicted for   $\sigma_{eff}$ appears  large compared to present data \cite{Aaboud:2018tiq}. Of course, what the above naive calculation might be telling us is that $\hat{f}({\bf q})$ is related not so much to the elastic but to an ``inelastic form factor''. Counting 4 protons being present in elastic events whereas only two (initial) protons being present in a true ''break up'' inelastic event, we expect only the second power and not the fourth power of the elastic form factor to appear in Eq.(\ref{g7}). If so,  
\begin{equation}
\label{g9}
\sigma_{eff}^{(inel)}(s) =\frac{1}{ \int[\frac{d^2{\bf q}}{(2\pi)^2}] \frac{1}{[1 + (q^2/t_o(s))]^4}}
= \Big{[} \frac{6\pi}{t_o(s)} \Big{]}; \ \ \ \ 
\sigma_{eff}^{inel}(13\ TeV) \approx\ 12.3\ milli-barns,
\end{equation}   
a bit closer to the phenomenological value estimated by  exploration of ATLAS compilation \cite{Aaboud:2018tiq}.\\
In this model the energy dependence of $\sigeff$ proceeds from that of the parameter $t_0(s)$. 
We notice here that in \cite{Fagundes:2013aja} we have shown that the presently available data for the differential elastic cross-section as well as the total cross-section, i.e.   the imaginary part of the forward elastic amplitude, can be described rather accurately through an expression which includes an energy dependent form factor. In this model,  $t_0(s)$ decreases with energy, hence  this model would predict $\sigeff(s) \uparrow \sqrt{s}$. We now turn to a discussion of the data and a comparison with the models we have just illustrated. 

\section{About  data and models}
\label{sec:datamodels}
\begin{figure}
\centering
\includegraphics[scale=0.26,angle=90]{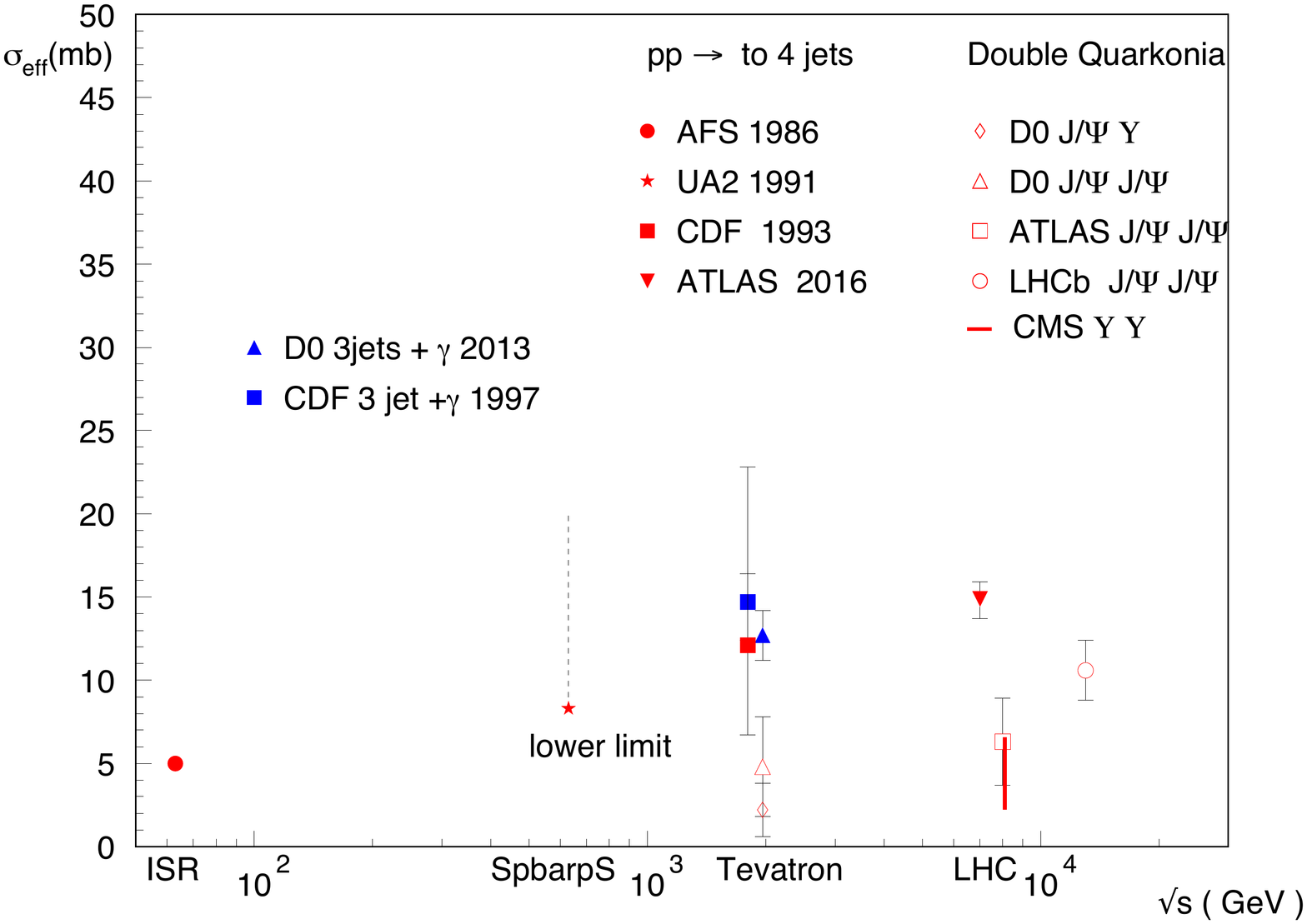}
\includegraphics[scale=0.26,angle=90]{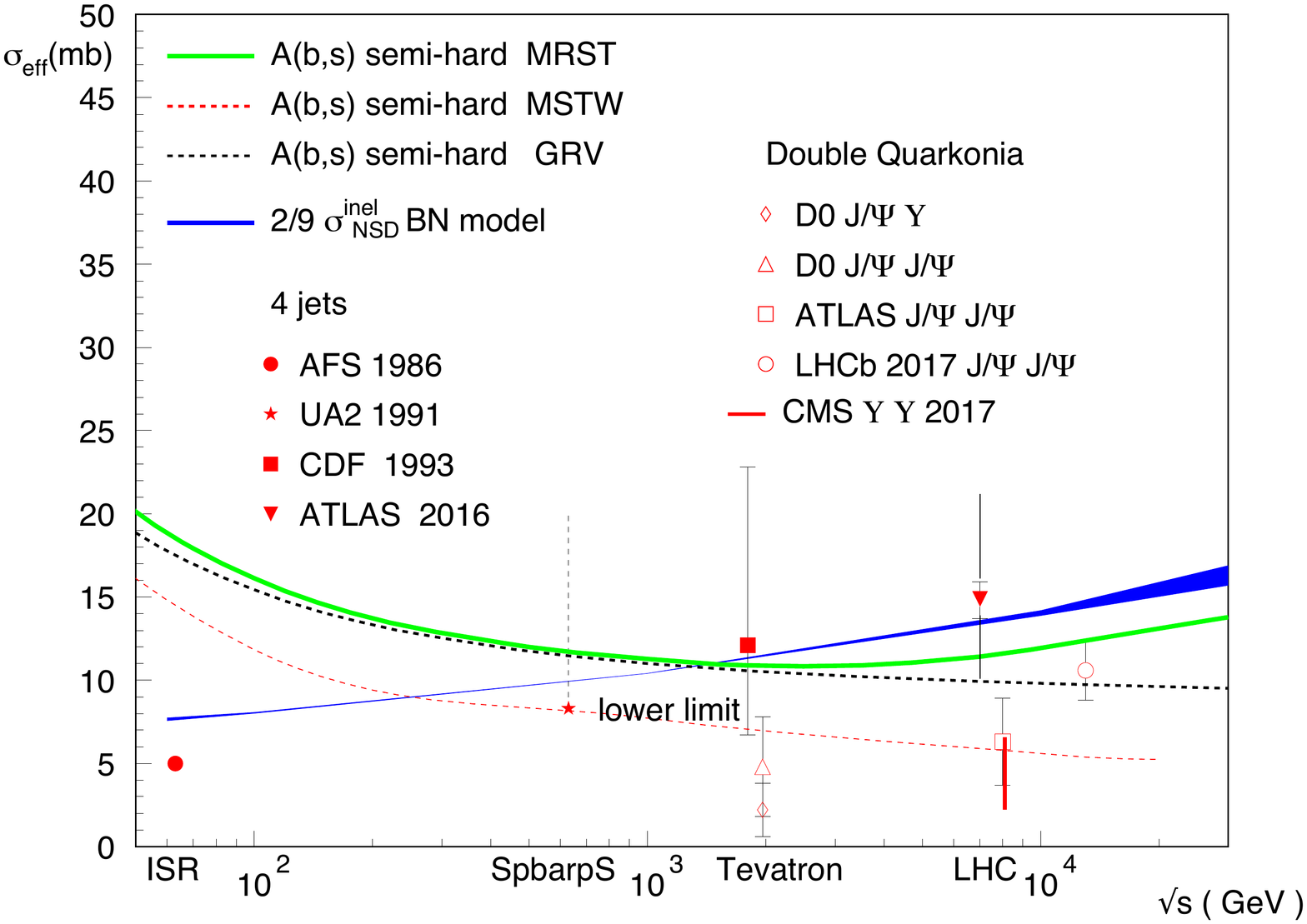}
\caption{ In the left panel,  existing DPS data as described in the text from Axial Field Spectrometer (AFS),
  \cite{Akesson:1986iv}, UA2 \cite{Alitti:1991rd}, CDF 1993 \cite{Abe:1993rv}  for ${\bar p} p \rightarrow 4 \  jets$ and  ATLAS at $\s=7$ TeV \cite{Aaboud:2016dea}
  for $ p p \rightarrow 4 \  jets$. We also have plotted  CDF 1997 \cite{Abe:1997xk} and D0 \cite{Abazov:2014fha} for $ {\bar p} p \rightarrow  \gamma \ 3 \  jets$, D0 \cite{Abazov:2014qba} for $ {\bar p} p \rightarrow J/\Psi \ J/\Psi $, ATLAS \cite{Aaboud:2016fzt} and LHCb for $  p p \rightarrow J/\Psi \ J/\Psi$ \cite{Aaij:2016bqq}, CMS for $p p \rightarrow \Upsilon \ \Upsilon$ \cite{Khachatryan:2016ydm}, D0 for ${\bar p} p \rightarrow J/\Psi \ \Upsilon$ \cite{Abazov:2015fbl}. At right, comparison of ${\bar p}/p \ p\rightarrow 4 \ jets$ or with ${\bar p}/p\ p\rightarrow quarkonia \ pair$ with two models described in the text. }
\label{fig:data-models}
\end{figure}

Available data not only span a very large energy range, but, as compiled by ATLAS, refer to  very different types of final states. This may indeed generate confusion since parton distributions, hence the calculation of $\sigeff$, differ according to  whether  the initial state be  mostly driven by gluon-gluon scattering or implicating valence quarks as well. Thus we have focused on similarly homogenous final states and show them in  the left panel  of Fig.~\ref{fig:data-models}.
The figure may indicate the following trends:
\begin{itemize}
\item for processes dominated by gluon gluon scattering, such as ${\bar p}/p\  p \rightarrow 4 jets$ and ${\bar p}/p \ p \rightarrow J/\Psi \ J/\Psi$, $\sigma_{eff}(\s) \uparrow \s$, although the scale is different, with  $\sigma_{eff}^{singlet}(\s) \approx \frac{1}{3}\sigma_{eff}(\s)^{all}$
\item for processes in which at least one of the final state particles must originate from a valence quark, as in $3 jets + \gamma$, the effective cross-section appears to be decreasing, as seen in the left  panel of Fig.\ref{fig:data-models} by the full blue symbols. 
\end{itemize}
In the right hand  panel we have compared the selected sets of data {\it vs.}  two models: the BN-inspired soft gluon resummation model described in Sect.~\ref{sec:BN}, 
and a  model based on the ansatz that all inclusive cross-sections rise. This model would be adequate to describe the case of gluon initiated  processes, less so when valence quarks initiate the process, as it is likely to be the case for the $3\ jets+ \gamma$ final state. Our ansatz, to describe $\sigeff$ for ${\bar p}/p \  p\rightarrow 4 \ jets,$
 is
 \begin{equation}
\sigeff\propto \siginel^{NSD}
\end{equation}  
We then use the  description of $\siginel^{NSD}$ from the  model of \cite{Fagundes:2013aja} and plot it as as blue band in the  right hand panel of Fig.~\ref{fig:data-models}, with an arbitrarily chosen factor $2/9$ for   normalization to the data.. We   consider  the two different cases of GRV or MSTW densities (MRST densities for total and inelastic cross-section are in good agreement with results from MSTW, as shown in the right hand panel Fig.~\ref{fig:qmax-sigjet-tot}).\\
 For the model which uses  $A(b)$ from soft gluon resummation, Sect.\ref{sec:BN}, we see that at LHC energies the model gives  good agreement with data, but the trend with energy is different. \\
 In summary for $pp\rightarrow 4\ jets$: 
\begin{itemize}
\item the impact parameter description as from Sect.~\ref{sec:BN} (green, red and dotted curves in Fig.~\ref{fig:data-models}) gives  an absolute   overall normalization of  LHC data     in   a good agreement with the plotted data, but is inconclusive as far as the energy dependence is concerned,
\item the scattering amplitude {\it cum} form factor model from Sect.~\ref{sec:ampli} would also reproduce the correct order of magnitude at LHC, and may indicate a rising $\sigeff$ from ISR to LHC, 
\item an empirical description from the NSD inelastic cross-section of \cite{Achilli:2011sw} would reproduce a rising energy trend from ISR to LHC.
\end{itemize}

\bibliographystyle{unsrt}
\bibliography{DPS}
\end{document}